\documentclass[12pt]{article}

\usepackage{amsmath}
\usepackage{amsfonts}
\usepackage{graphicx}
\usepackage{caption}
\usepackage{subcaption}

\usepackage{authblk}
\usepackage{amssymb}

\usepackage{epstopdf}
\usepackage{epsfig}
\usepackage{hyperref}

\usepackage{color}

\usepackage{float}

\usepackage[normalem]{ulem}

\date{}

\begin{document}

	\title{Radial perturbations of scalar-Gauss-Bonnet black holes beyond spontaneous scalarization}
	
	\author[1]{Jose Luis Bl\'azquez-Salcedo \thanks{\href{mailto:jose.blazquez.salcedo@uni-oldenburg.de}{jlblaz01@ucm.es}}}
	
	\author[2,3]{Daniela D. Doneva
		\thanks{\href{mailto:daniela.doneva@uni-tuebingen.de }{daniela.doneva@uni-tuebingen.de }}}
	
	\author[1]{Jutta Kunz \thanks{\href{mailto:jutta.kunz@uni-oldenburg.de}{jutta.kunz@uni-oldenburg.de}}} 
	
	\author[2,4,5]{Stoytcho S. Yazadjiev \thanks{\href{mailto:yazad@phys.uni-sofia.bg}{yazad@phys.uni-sofia.bg}}}
	
	\affil[1]{Institut f\"ur  Physik, Universit\"at Oldenburg, Postfach 2503,
		D-26111 Oldenburg, Germany}
	\affil[2]{Theoretical Astrophysics, Eberhard Karls University of T\"ubingen, T\"ubingen 72076, Germany}
	\affil[3]{INRNE - Bulgarian Academy of Sciences, 1784  Sofia, Bulgaria}
	\affil[4]{Department of Theoretical Physics, Faculty of Physics, Sofia University, Sofia 1164, Bulgaria}
	\affil[5]{Institute of Mathematics and Informatics, Bulgarian Academy of Sciences, Acad. G. Bonchev Street 8, Sofia 1113, Bulgaria}

	\maketitle
	
	\begin{abstract}
		Spontaneous scalarization of black holes in scalar-Gauss-Bonnet (sGB) gravity is a very interesting phenomenon allowing black holes to circumvent the no-hair theorem and acquire scalar hair while leaving the weak field regime of the theory practically unaltered. It was recently shown that if we allow for a different form of the coupling between the scalar field and the Gauss-Bonnet invariant, a new type of scalarization is possible beyond the standard one that can be excited only nonlinearly. In this case the spectrum of black hole solutions can be more complicated, naturally opening the question about their stability. The goal of the present paper is to study this question including the possible loss of hyperbolicity of the radial perturbation equation. We show that one of the nonlinearly scalarized phases of the Schwarzschild black hole can be stable and hyperbolic for large enough black hole masses. These results establish the studied hairy black holes as viable astrophysical candidates. 
	\end{abstract}
	
	\section{Introduction}
	Testing the nature of the black holes has been an important goal for decades. There is a plethora of observations that pave the way towards determining whether the spacetime around such compact objects in indeed described by the Kerr metric \cite{Berti:2015itd}. Especially with the advance of the gravitational wave astronomy, as well as the X-ray observations and the observations of black hole shadows, we are closer than ever to put strong limits on the deviations from general relativity (GR) or even more interesting -- to observe a phenomenon that does not fit well in the standard picture and is a signal for a modification of Einstein's gravity. The advance in observations calls for a further advance in theory. That is why it is very important to understand as well as possible the deviations from the Schwarzschild or Kerr metric that different alternative theories of gravity offer and especially their astrophysical implications. Even though this topic has advanced a lot in the last decades there is still a lot to be done. One of the reasons is that for a large class of modified theories of gravity there are uniqueness theorems stating that the solutions should be the same as in GR. Thus, it is not always easy to find a theory predicting a viable alternative to the Kerr solutions while still being in agreement with the weak field observations where GR is tested with remarkable accuracy \cite{Will:2005va,Faraoni:2010pgm,Berti:2015itd,CANTATA:2021mgk}.
	
	Among the interesting theories that fall into this category are the extended scalar-tensor theories allowing for the development of the so-called  scalarization. Their essence is that they are perturbatively equivalent to GR for weak fields while second order phase transitions to a scalarized state of the compact object can occur for strong fields that is called spontaneous scalarization \cite{Damour:1993hw,Damour:1996ke,Doneva:2017bvd,Silva:2017uqg,Antoniou:2017acq}. Even though initially this idea was developed in the framework of neutron stars \cite{Damour:1993hw,Damour:1996ke}, it was recently found that in certain types of modified theories of gravity the spacetime curvature itself can act as a source of the scalar field. This was first discovered in the scalar-Gauss-Bonnet (sGB) theory \cite{Doneva:2017bvd,Silva:2017uqg,Antoniou:2017acq,Cunha:2019dwb,Collodel:2019kkx,Dima:2020yac,Herdeiro:2020wei,Berti:2020kgk} but later it was demonstrated that a much larger class of modified theories of gravity admits such scalarization \cite{Herdeiro:2018wub,Andreou:2019ikc,Gao:2018acg,Doneva:2021dcc}. The dynamics of these black holes was also addressed in a series of papers \cite{Witek:2018dmd,Witek:2020uzz,Silva:2020omi,Doneva:2021dqn,Kuan:2021lol,East:2021bqk}. Further studies in sGB gravity have focused on the linear stability of these spontaneously scalarized black holes \cite{Blazquez-Salcedo:2018jnn,Blazquez-Salcedo:2020rhf,Blazquez-Salcedo:2020caw} and on interesting effects that one can have if the coupling function between the scalar field and the Gauss-Bonnet invariant is modified or a potential is included for the scalar field \cite{Minamitsuji:2018xde,Silva:2018qhn,Doneva:2019vuh,Macedo:2019sem,Bakopoulos:2020dfg}. All these studies considered the case of standard scalarization when we have the Schwarzschild (or Kerr) black hole being destabilized below a certain black hole mass and the scalarized solutions branching out at that point. Thus any initially arbitrarily small perturbation of the Schwarzschild black hole, when in the unstable range, will result in an exponential growth of the scalar field that will be quenched at a certain point by nonlinear mechanisms forming an equilibrium black hole configuration endowed by scalar hair. 
	
	For certain scalar field couplings, though, another very interesting effect can be observed that we call fully nonlinear scalarization. That is when the Schwarzschild black hole is always linearly stable but still black holes with nonzero scalar hair exist that are energetically more favourable \cite{Doneva:2021tvn}. An analogous phenomenon has been observed in the case of the charged black holes \cite{Blazquez-Salcedo:2020nhs,LuisBlazquez-Salcedo:2020rqp,Blazquez-Salcedo:2020crd}. We call them scalarized phases of the Schwarzschild black hole and they can be excited only if a strong enough perturbation is imposed. It was found in \cite{Doneva:2021tvn} that the spectrum of hairy black hole solutions can be complicated and a natural question that arises is about their stability. This is the topic of the present paper where we study the radial stability of the scalarized phases with the idea to identify the stable and hyperbolic parts of the branches that will be also of interest to astrophysics. 
	
	The structure of the paper is as follows. In Section \ref{sec:theory_and_ansatz} we present the basics of the employed sGB theory and the construction of the background black hole solutions. Section \ref{sec:Radial_Perturbations} is devoted to the presentation of the approach we follow in order to perturb the field equations and solve the radial perturbations problem. The main results of the paper about the stability and hyperbolicity of the scalarized phases are presented in \ref{sec:StabilityAnalysis}. The paper ends with Conclusions.

	\section{Theory and ansatz} \label{sec:theory_and_ansatz}
	The general form of the action in sGB gravity can be written in the following way
	\begin{eqnarray}
		S=&&\frac{1}{16\pi}\int d^4x \sqrt{-g} 
		\Big[R - 2\nabla_\mu \varphi \nabla^\mu \varphi 
		+ \lambda^2 f(\varphi){\cal R}^2_{GB} \Big] ,\label{eq:quadratic}
	\end{eqnarray}
	where $R$ is the Ricci scalar with respect to the spacetime metric $g_{\mu\nu}$ and $\varphi$ is the scalar field. The  Gauss-Bonnet invariant entering the action ${\cal R}^2_{GB}$ is defined as ${\cal R}^2_{GB}=R^2 - 4 R_{\mu\nu} R^{\mu\nu} + R_{\mu\nu\alpha\beta}R^{\mu\nu\alpha\beta}$ where $R_{\mu\nu}$ is the Ricci tensor and $R_{\mu\nu\alpha\beta}$ is the Riemann tensor. $f(\varphi)$ is the coupling function between the scalar field and ${\cal R}^2_{GB}$, while  $\lambda$ is the Gauss-Bonnet coupling constant that has dimension of $length$.

	We will consider static and spherically symmetric spacetimes as well as static and spherically symmetric scalar field configurations. Thus the ansatz for the spacetime metric can be taken as
	\begin{eqnarray}
		ds^2= - e^{2\Phi(r)}dt^2 + e^{2\Lambda(r)} dr^2 + r^2 (d\theta^2 + \sin^2\theta d\phi^2 ). 
	\end{eqnarray}   
	Using it we can derive the dimensionally reduced field equations that will describe the spacetime around the unperturbed black holes
	\begin{eqnarray}
		&&\frac{2}{r}\left[1 +  \frac{2}{r} (1-3e^{-2\Lambda})  \Psi_{r}  \right]  \frac{d\Lambda}{dr} + \frac{(e^{2\Lambda}-1)}{r^2} 
		- \frac{4}{r^2}(1-e^{-2\Lambda}) \frac{d\Psi_{r}}{dr} - \left( \frac{d\varphi}{dr}\right)^2=0, \label{DRFE1}\\ && \nonumber \\
		&&\frac{2}{r}\left[1 +  \frac{2}{r} (1-3e^{-2\Lambda})  \Psi_{r}  \right]  \frac{d\Phi}{dr} - \frac{(e^{2\Lambda}-1)}{r^2} - \left( \frac{d\varphi}{dr}\right)^2=0,\label{DRFE2}\\ && \nonumber \\
		&& \frac{d^2\Phi}{dr^2} + \left(\frac{d\Phi}{dr} + \frac{1}{r}\right)\left(\frac{d\Phi}{dr} - \frac{d\Lambda}{dr}\right)  + \frac{4e^{-2\Lambda}}{r}\left[3\frac{d\Phi}{dr}\frac{d\Lambda}{dr} - \frac{d^2\Phi}{dr^2} - \left(\frac{d\Phi}{dr}\right)^2 \right]\Psi_{r} 
		\nonumber \\ 
		&& \hspace{0.9cm} - \frac{4e^{-2\Lambda}}{r}\frac{d\Phi}{dr} \frac{d\Psi_r}{dr} + \left(\frac{d\varphi}{dr}\right)^2=0, \label{DRFE3}\\ && \nonumber \\
	\end{eqnarray}
	\begin{eqnarray}
		&& \frac{d^2\varphi}{dr^2}  + \left(\frac{d\Phi}{dr} \nonumber - \frac{d\Lambda}{dr} + \frac{2}{r}\right)\frac{d\varphi}{dr} \nonumber \\ 
		&& \hspace{0.9cm} - \frac{2\lambda^2}{r^2} \frac{df(\varphi)}{d\varphi}\Big\{(1-e^{-2\Lambda})\left[\frac{d^2\Phi}{dr^2} + \frac{d\Phi}{dr} \left(\frac{d\Phi}{dr} - \frac{d\Lambda}{dr}\right)\right]    + 2e^{-2\Lambda}\frac{d\Phi}{dr} \frac{d\Lambda}{dr}\Big\} =0, \label{DRFE4}
	\end{eqnarray}
	where
	\begin{eqnarray}
		\Psi_{r}=\lambda^2 \frac{df(\varphi)}{d\varphi} \frac{d\varphi}{dr}.
	\end{eqnarray}
	We further assume that the cosmological value of the scalar field vanishes $\varphi|_{r\rightarrow\infty}\equiv\varphi_{\infty}=0$,

	We will focus our study on three different coupling functions all of them leading to nonlinearly scalarized black hole solutions in the spirit of \cite{Doneva:2021tvn}:
	\begin{equation}
		f_I(\varphi)=  \frac{1}{4\kappa}\left(1-e^{-\kappa \varphi^4}\right), \label{eq:coupling_function_I}
	\end{equation}
	\begin{equation}
		f_{II}(\varphi)=  \frac{1}{6\kappa}\left(1-e^{-\kappa \varphi^6}\right), \label{eq:coupling_function_II}
	\end{equation}
	\begin{equation}
		f_{III}(\varphi)=  \frac{1}{2\beta}\left(1-e^{-\beta (\varphi^2+\kappa\varphi^4)}\right). \label{eq:coupling_function_III}
	\end{equation}
	All the coupling functions are constructed so that the conditions $f(0)=0$ and $df/d\varphi(0)=0$ are fulfilled. The former condition can be imposed since the field equations include only the first derivative of the coupling function but not the coupling function itself, while the latter condition guarantees that the Schwarzschild solution is also a solution to the sGB field equations with $\varphi=0$. The first two couplings with a pure quartic and sextic scalar field term in the exponent, though, have an important difference compared to the third one since they satisfy different conditions on the second derivative with respect to the scalar field. More specifically,
	\begin{eqnarray}
		\frac{d^2f_{(I\;{\rm or}\;II)}}{d\varphi^2}(0)=0,\;\; {\rm while}\;\;\ \frac{d^2f_{III}}{d\varphi^2}(0)>0. \label{eq:CouplingFuncConditions}
	\end{eqnarray}
	The first condition for $f_I$ and $f_{II}$ leads to the fact that no tachyonic instability is possible and the Schwarzschild solution is always stable against linear scalar perturbations. Thus if black holes with scalar hair exist with such type of coupling, they should form  a new black hole phase coexisting with the stable Schwarzschild black hole phase that we will call a scalarized black hole phase. 
	
	The second condition in \eqref{eq:CouplingFuncConditions} for $f_{III}$ leads to a tachyonic destabilization of the Schwarzschild black hole below a certain mass that is the mechanism for the appearance of the standard spontaneously scalarized black holes \cite{Doneva:2017bvd,Silva:2017uqg,Antoniou:2017acq}. The quartic scalar field term in $f_{III}(\varphi)$, though, changes the spectrum of the solutions considerably and for certain combinations of parameters the co-existence of both nonlinearly and standard scalarized black holes is allowed. We refer the reader to \cite{Doneva:2021tvn} for a detailed discussion of all these types of scalarization.
	
	In order to construct the black hole solutions we have to impose the following conditions at the black hole horizon $r=r_H$ 
	\begin{eqnarray}
		e^{2\Phi}|_{r\rightarrow r_H} \rightarrow 0, \;\; e^{-2\Lambda}|_{r\rightarrow r_H} \rightarrow 0. \label{eq:BC_rh}
	\end{eqnarray} 
	If one requires regularity of the metric functions and the scalar field then the following condition should be also satisfied
	\begin{equation}
		r_H^4 > 24 \lambda^4 \left(\frac{df}{d\varphi}(\varphi_{H})\right)^2, \label{eq:BC_sqrt_rh}
	\end{equation}
	where $\varphi_{H}$ is the value of the scalar field at the horizon. In the calculations below we will use the following natural definition of a  singular limit described by the curve $(r_H/\lambda)^4 - 24 \left(\frac{df}{d\varphi}(\varphi_{H})\right)^2=0$ in the $(\varphi_{H},r_H/\lambda)$ plane, that separates the region of the possible existence of hairy black holes from the region where the regularity condition is violated, and only the Schwarzschild black hole can be a solution of the field equations.
	
	At infinity asymptotic flatness means
	\begin{eqnarray}
		\Phi|_{r\rightarrow\infty} \rightarrow 0, \;\;  \Lambda|_{r\rightarrow\infty} \rightarrow 0,\;\; \varphi|_{r\rightarrow\infty} \rightarrow 0\;\;.   \label{eq:BH_inf}
	\end{eqnarray} 
	The leading order scalar field asymptotic for the black hole solutions we will be interested in, assuming zero background scalar field at infinity, has the form
	\begin{equation}
		\varphi|_{r\rightarrow\infty} \sim \frac{D}{r},
	\end{equation}
	where the constant $D$ denotes the scalar charge of the back hole solutions.
	
	\section{Radial Perturbations}\label{sec:Radial_Perturbations}
	
	\subsection{Ansatz and equations}
	
	Having discussed the equations that determine the background black hole solutions, let us now turn to considering their radial perturbations with the end goal to derive the master equation governing these perturbations. In what follows, we adopt the same procedure as in \cite{Blazquez-Salcedo:2018jnn}. Namely, we assume the following form of the metric and scalar field perturbations:
	\begin{eqnarray} 
		\label{metric_pert}
		ds^2 &=& -e^{2\Phi(r)+\epsilon F_t(r,t)}dt^2 + e^{2\Lambda(r) + \epsilon F_r(r,t)} dr^2 + r^2 (d\theta^2 + \sin^2\theta d\phi^2), \nonumber \\
		\varphi &=& \varphi_0(r) + \epsilon \varphi_1(r,t),
	\end{eqnarray}
	with $\epsilon$ being the control parameter of the perturbations. By writing the field equations in terms of this metric and keeping only first order terms in $\epsilon$, one can derive a system of perturbation equations for the metric and scalar field perturbations $F_t(r,t)$, $F_r(r,t)$ and $\varphi_1(r,t)$. This system can be simplified into a single second order differential equation of the form
	\begin{equation}
		\label{wave_eq}
		g^2(r)  \frac{\partial^2\varphi_1}{\partial t^2}   - \frac{\partial^2\varphi_1}{\partial r^2} + C_1(r) \frac{\partial\varphi_1}{\partial r}  
		+  U(r)  \varphi_1=0,
	\end{equation}
	where the functions $U(r)$, $g(r)$ and $C_1(r)$ depend only  on the background metric and scalar field, with complicated expressions that can be found in \cite{Blazquez-Salcedo:2018jnn}.
	
	In order to study the mode stability of the background configuration, we decompose the perturbation function $\varphi_1$ as
	\begin{equation}
		\varphi_1(r,t) = \varphi_1(r) e^{i\omega t},
	\end{equation}
	where $\omega$ is in general a complex number.  
	{We obtain the master equation for the eigenvalue problem, namely}
	\begin{eqnarray}
		\label{master_eq} 
		\frac{d^2\varphi_1}{dr^2} = C_1(r) \frac{d\varphi_1}{dr} + \left[ U(r) - \omega^2 g^2(r) \right] \varphi_1(r).
	\end{eqnarray} 
	
	{The master equation (\ref{master_eq}) can be cast into the standard Schr\"odinger form by defining the function $Z(r)$:}
	\begin{eqnarray}
		\varphi_1(r) = C_0(r) Z(r), 
	\end{eqnarray}
	{where $C_0(r)$ is the solution of the following differential equation}
	\begin{equation}
		\frac{1}{C_0}\frac{dC_0}{dr}= C_1 - \frac{1}{g}\frac{dg}{dr}.
	\end{equation}
	{Thus we obtain}
	\begin{eqnarray}
		\label{master_eq_simp} 
		\frac{d^2Z}{dR^2} = \left[ V(R) - \omega^2 \right] Z,
	\end{eqnarray} 
	where we have defined the tortoise coordinate $R$ and the effective potential as
	\begin{eqnarray}
		\frac{dR}{dr} &=& g, \label{eq:g}\\
		V(R) &=&  \frac{1}{g^2} \left( U + \frac{C_1}{C_0}\frac{dC_0}{dr}-\frac{1}{C_0}\frac{d^2C_0}{dr^2} \right).
		\label{eq:potential}
	\end{eqnarray}

	\subsection{Boundary conditions and numerical method}
	
	We want the perturbation to have the form of an outgoing wave at infinity and an ingoing wave at the horizon. Since we are interested in the stability analysis of the background solutions, we will focus on perturbations with purely imaginary eigenfrequencies: $\omega = i\omega_I$. In such a case the asymptotic behaviour of the perturbation function $Z$ becomes:
	
	\begin{eqnarray}
		Z \xrightarrow[r \to \infty]{} e^{i\omega(t-R)} = e^{-\omega_I (t-R)}, \\
		Z \xrightarrow[r \to r_H]{} e^{i\omega(t+R)} = e^{-\omega_I (t+R)}.
	\end{eqnarray}

	The asymptotic behaviour of unstable perturbations possessing $\omega_I < 0$ simplifies considerably, and it is straightforward to show that the function $Z$ must satisfy the following boundary conditions:
	
	\begin{eqnarray}
		Z|_{r=\infty}=Z|_{r=r_H}=0.  \label{bc:Z}
	\end{eqnarray}

	Our main goal in this paper is to check the desired black hole solutions for stability using the radial perturbation equation \eqref{master_eq_simp}. In order to do so we solve this equation together with the boundary conditions (\ref{bc:Z}), and calculate numerically the unstable modes with $\omega_I<0$, if they exist. The numerical procedure is the same as in \cite{Blazquez-Salcedo:2018jnn}, and we refer there for more details. The free parameters we have in our system are $r_H$, $\lambda$ and the coupling constant $\kappa$ (and $\beta$ in the case of $f_{III}$). The numerical procedure allows us to explore the dependence of the unstable mode on these parameters.

	A lot of information can be gained, though, only on the basis of the behaviour of the potential $V(R)$ in eq.~\eqref{eq:potential} and more precisely on the sign of the integral
	\begin{eqnarray}
		\int^{\infty}_{-\infty} V(R) dR \ . \label{int:V}
	\end{eqnarray}
	A negative value is a signal for the instability of the black hole solutions. That is why, we monitor the sign of the potential and the value of this integral as we explore the domain of existence of the solutions.

	Another important property of the perturbation equation is its hyperbolicity. As well known, the time-dependent field equations in Gauss-Bonnet theory suffers from loss of hyperbolicity both in their linear and nonlinear form \cite{Blazquez-Salcedo:2018jnn,Ripley:2019hxt,Blazquez-Salcedo:2020rhf,East:2021bqk,Kuan:2021lol}. For the standard scalarization this normally happens for small mass black holes. It is interesting to observe whether such hyperbolicity loss is also present for the scalarized black hole phases with the coupling functions listed above. In order to explore this, we study the positivity of the function $(1-\frac{r_H}{r})^2 g^2$. We say that hyperbolicity of a configuration is lost if this function takes zero or negative values. Hence we also monitor the behaviour of this function as we vary the free parameters of our system.
	
	At the end let us point out that what we consider in the present paper is only the loss of hyperbolicity with respect to the radial perturbation equations. As demonstrated in \cite{Blazquez-Salcedo:2020rhf} the black hole mass where the evolution equation governing the axial perturbations changes its character can be a bit different.  The two points for the radial and axial perturbations are very close, though, at least for the case of the standard scalarization \cite{Blazquez-Salcedo:2018jnn,Blazquez-Salcedo:2020rhf}, and that is why we expect that the conclusions made in the present paper will not be significantly altered even if other types of perturbations are considered. 
	
	\section{Stability analysis}\label{sec:StabilityAnalysis}
	
	Now we start analysing the stability of the black hole solutions separately for each of the three coupling functions \eqref{eq:coupling_function_I}--\eqref{eq:coupling_function_III}. We will comment in detail on the spectrum of solutions and their domain of existence, the stability properties and the possible loss of hyperbolicity of the radial perturbations equation. 
	
	\subsection{Results for coupling $f_I(\varphi)=  \frac{1}{4\kappa}\left(1-e^{-\kappa \varphi^4}\right)$}
	This is the perhaps simplest type of coupling function that can lead to the existence of stable scalarized black hole phases of the stable Schwarzschild solution. Indeed, a similar phenomenon might appear for simpler types of coupling when $f(\varphi)$ is proportional to $\varphi^4$, but our studies in \cite{Doneva:2021tvn} have shown that even if scalarized phases exist in this case, they are always unstable and thus not of interest. Other studies of coupling functions, containing some type of $\varphi^4$ terms, without considering the phenomenon of nonlinear scalarized phases, though, were performed e.g. in \cite{Antoniou:2017hxj,Minamitsuji:2018xde,Silva:2018qhn,Corelli2020}.
	
	Let us start with a brief description of the domain of existence and the structure of scalarized black hole branches. It changes with the value of $\kappa$, but essentially seems to fall into one of three possible cases. 
	
	\begin{figure}[H]
		\centering
		\includegraphics[width=0.38\textwidth,angle=-90]{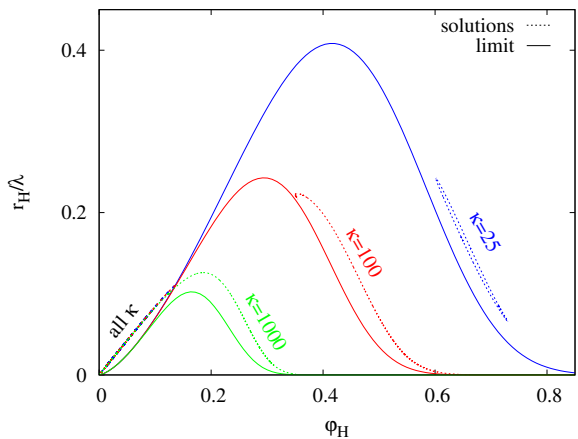}
		\caption{Domain of existence of black holes parametrized by $r_H/\lambda$ and $\varphi_H$ for the first coupling $f_I$ and different values of $\kappa$ (blue, red and green for $\kappa=25$, $100$ and $1000$ respectively). Dashed lines correspond to black hole solutions while the solid line represents the singular limit curve.}
		\label{Fig:domain_I}
	\end{figure}

	In figure \ref{Fig:domain_I} we show $r_H/\lambda$ vs $\varphi_H$ for the first coupling $f_I$ eq.~\eqref{eq:coupling_function_I} and $\kappa=25,100,1000$ in blue, red and green, respectively, that represents small, intermediate and large values of $\kappa$. The solid lines correspond to the singular limit curve that is determined by eq.~\eqref{eq:BC_sqrt_rh} and separates the region where black hole solutions with scalar hair can exist from the region where their existence is not possible due to violation of the regularity condition at the black hole horizon. The dashed lines represent the scalarized black hole phases.
	
	For $\kappa=1000$ (green), all solutions are smoothly connected, and the solution curve lies above the limit curve for this value of $\kappa$. As one decreases $\kappa$, the maximum of the limit curve rises and eventually intersects the curve of solutions, dividing it in two pieces. This is the case for $\kappa=100$ (red) where it can be seen that the solutions are separated in two different sets, one to the left of the limit curve, and another one to the right. Interestingly, as one further decreases $\kappa$, the right part of the curve of solutions deforms into a closed loop. This is the case of $\kappa=25$ (blue). Remember that for every value of $\kappa$, including the closed loop case, we have a branch of hairy black hole solutions spanning from zero black hole mass to the intersection with the line of existence. Due to resolution reasons, though, it is difficult to distinguish these branches for different $\kappa$ in Fig.~\ref{Fig:domain_I} (see for instance how the solutions for all $\kappa$ are superimposed on the bottom left corner of this figure).  Below we will consider the stability of each of these cases separately. Let us just summarize in advance that the results indicate that only large and intermediate values of $\kappa$ possess both radially stable and hyperbolic solutions. In addition, all conclusions regarding radial stability are in agreement with the thermodynamic analysis performed in \cite{Doneva:2021tvn}.

	\subsubsection{Coupling  $f_{I}$, large $\kappa=1000$}
	
	\begin{figure}[H]
		\centering
		\includegraphics[width=0.38\textwidth,angle=-90]{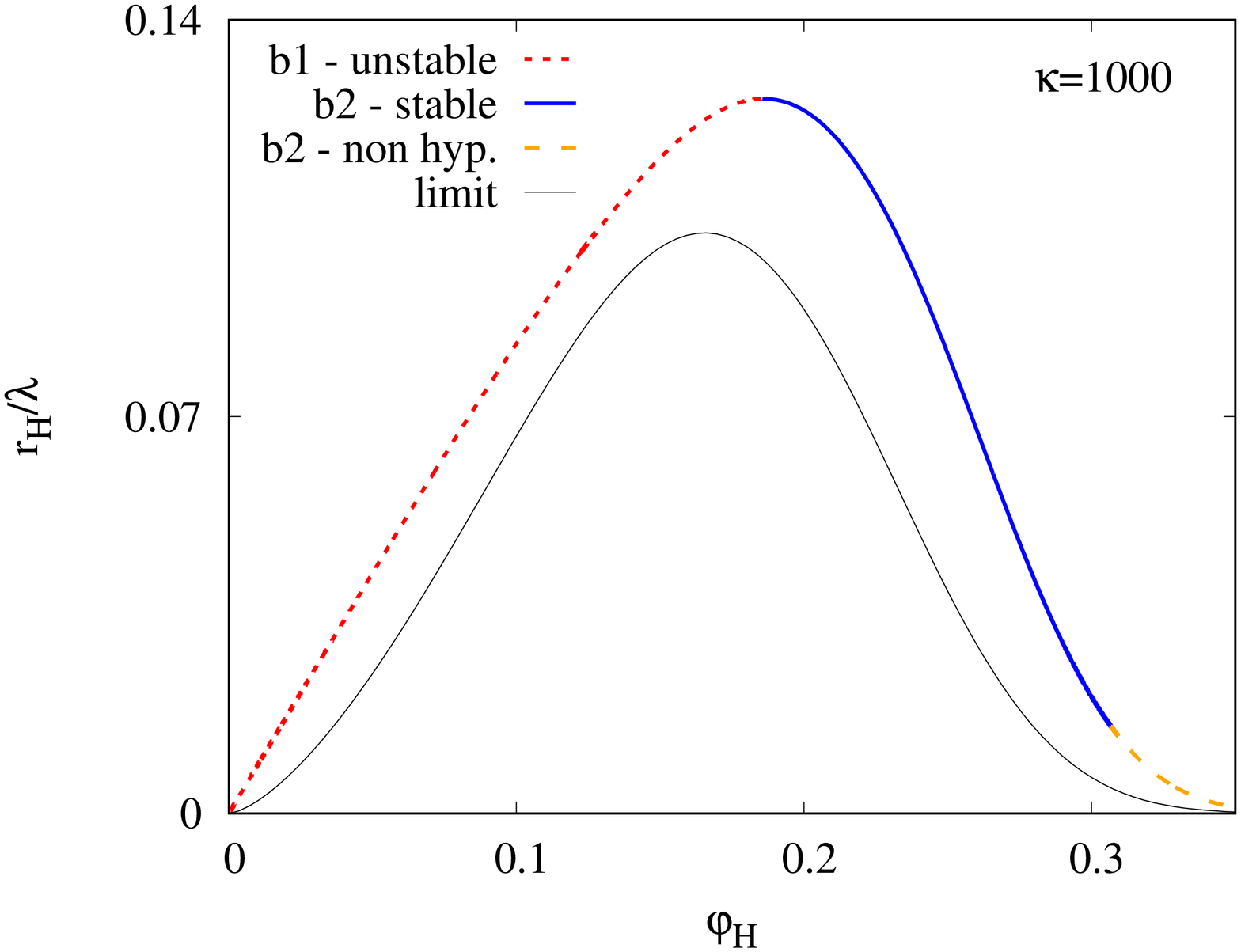}
		\includegraphics[width=0.38\textwidth,angle=-90]{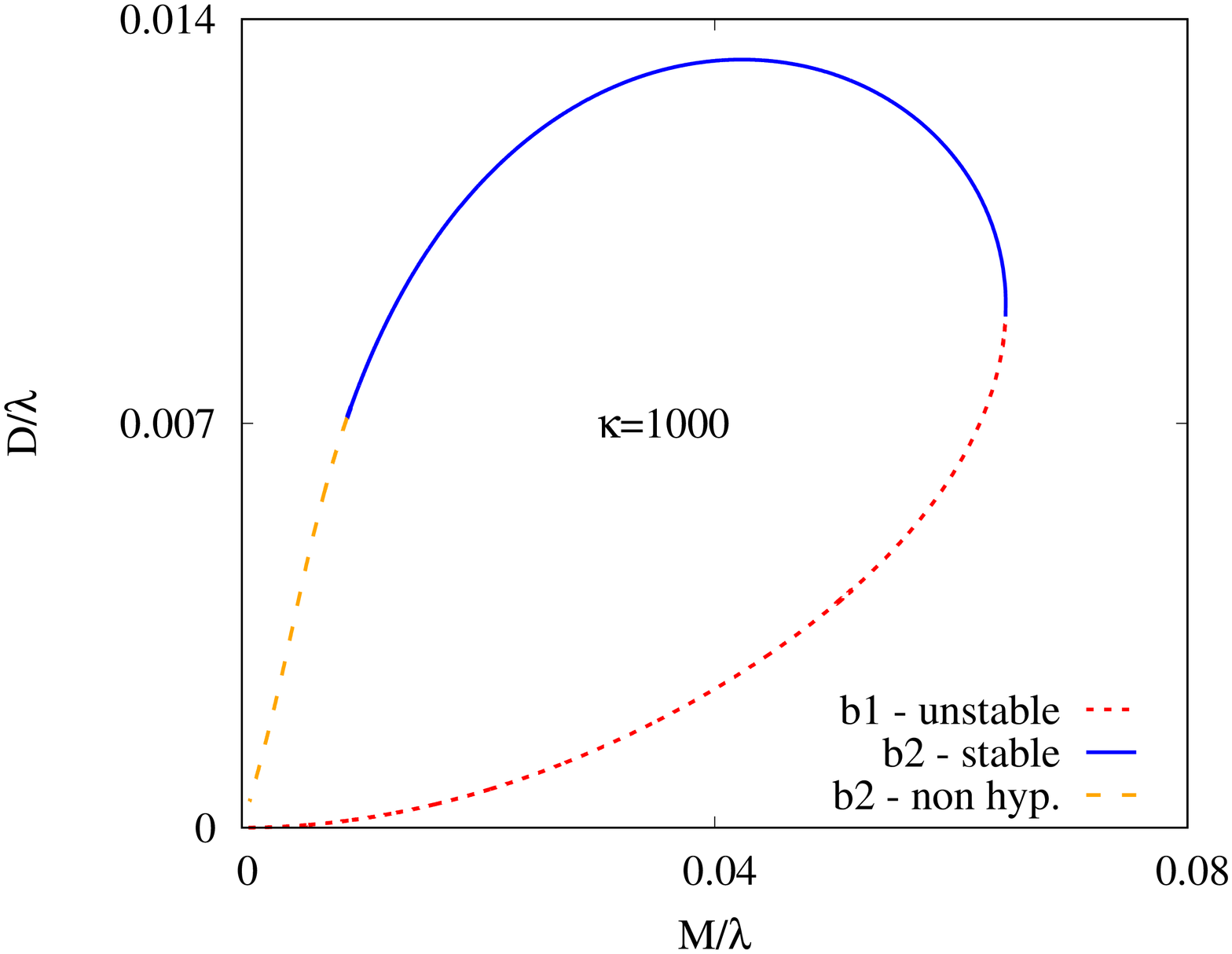}
		\includegraphics[width=0.38\textwidth,angle=-90]{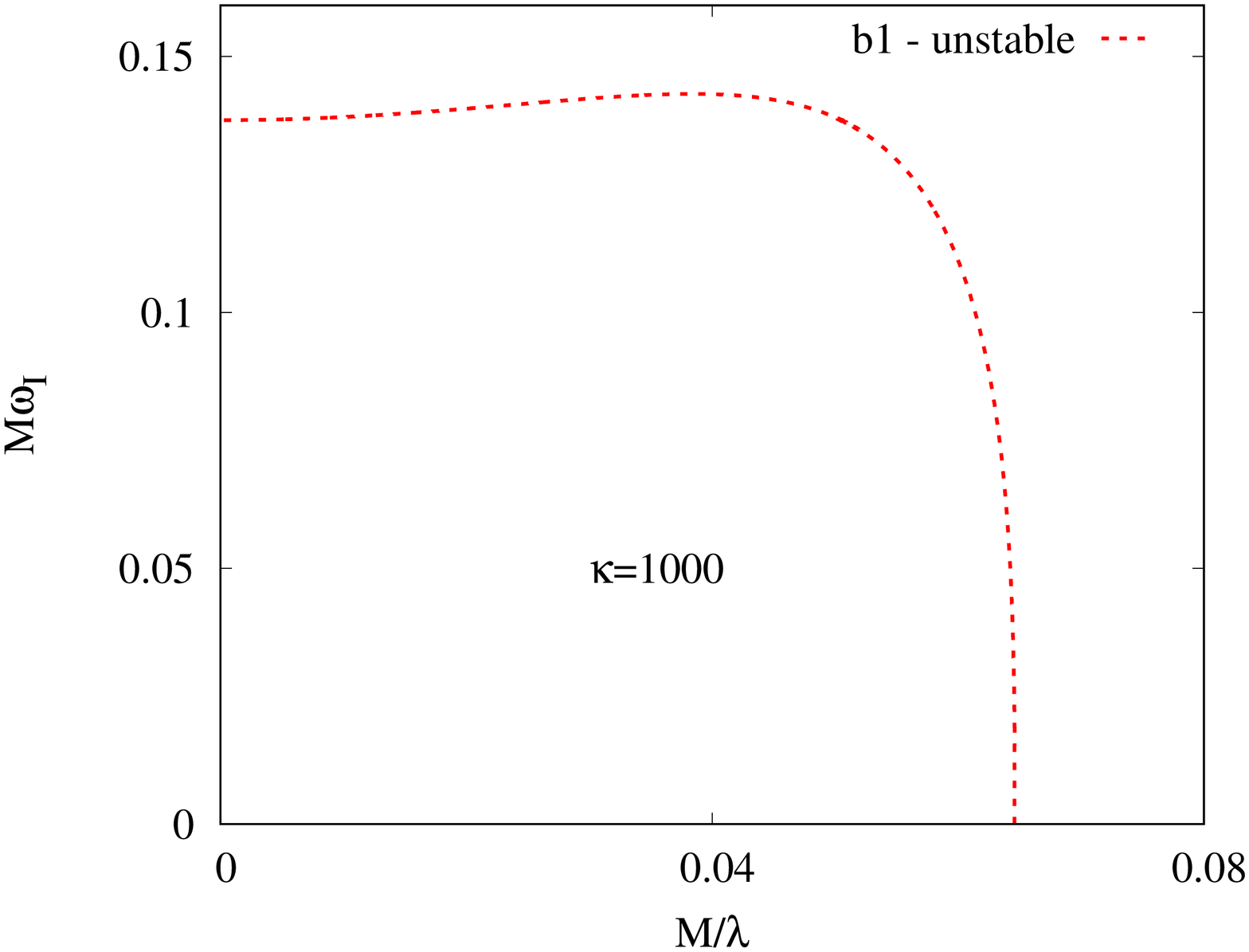}
		\caption{The results for coupling $f_I$ and large $\kappa=1000$ are plotted. \textit{(top-left panel)} The normalized black hole radius  $r_H/\lambda$  as a function of $\varphi_H$. \textit{(top-right panel)} The normalized black hole scalar charge  $D/\lambda$  as a function of the black hole mass $M/\lambda$. The solid black line represents the singular limit curve while the lines with different colors correspond to different branches of hairy black hole. \textit{(bottom panel)} The imaginary part of the frequency normalized with the black holes mass $M \omega_I$ for the only unstable b1 branch as a function of $M/\lambda$.}
		\label{Fig:k1000_I}
	\end{figure}
	
	In figure \ref{Fig:k1000_I} we plot the results for $f_I$ and $\kappa=1000$ that represent the large $\kappa$ case discussed above. In the top-left panel we show again $r_H/\lambda$ vs $\varphi_H$ but with a clear separation between the different parts of the branch that possess different properties such as (in)stability and (loss of) hyperbolicity. We have introduced the following naming convention of the branches: We call b1 (in red) the part of the branch that spans from zero black hole mass until the maximum of $r_H$ while b2 (in blue and orange) is the second part of the sequence after this maximum. Additional information about the black hole branches can be gained from the top-right panel of figure \ref{Fig:k1000_I} where we show the scalar charge $D/\lambda$ vs $M/\lambda$. It can be seen there that the solutions in b1 are always below the solutions in b2 and thus have a lower scalar charge.
	
	Solutions in b1 (red curve) are always unstable since the integral of the potential (\ref{int:V}) is negative and it is possible to find an unstable mode for all these configuration. This can be seen in the bottom panel of figure \ref{Fig:k1000_I} where the imaginary part of the frequency $\omega_I$ for the b1 branch is plotted as a function of $M/\lambda$ \footnote{Remember that for unstable modes the real part of the frequency is always zero.}. The black hole with the maximum value of $M/\lambda$ possesses a zero mode.
	
	The solutions belonging to the b2 branch are stable. However, for small enough values of $M/\lambda$, the radial perturbations become non-hyperbolic (orange curve in the top panels in Fig. \ref{Fig:k1000_I}). This non-hyperbolic part is smaller for larger $\kappa$ while it can increase significantly with the decrease of $\kappa$ as we will see below.

	\subsubsection{Coupling  $f_{I}$, intermediate $\kappa=100$}
	\begin{figure}[H]
		\centering
		\includegraphics[width=0.35\textwidth,angle=-90]{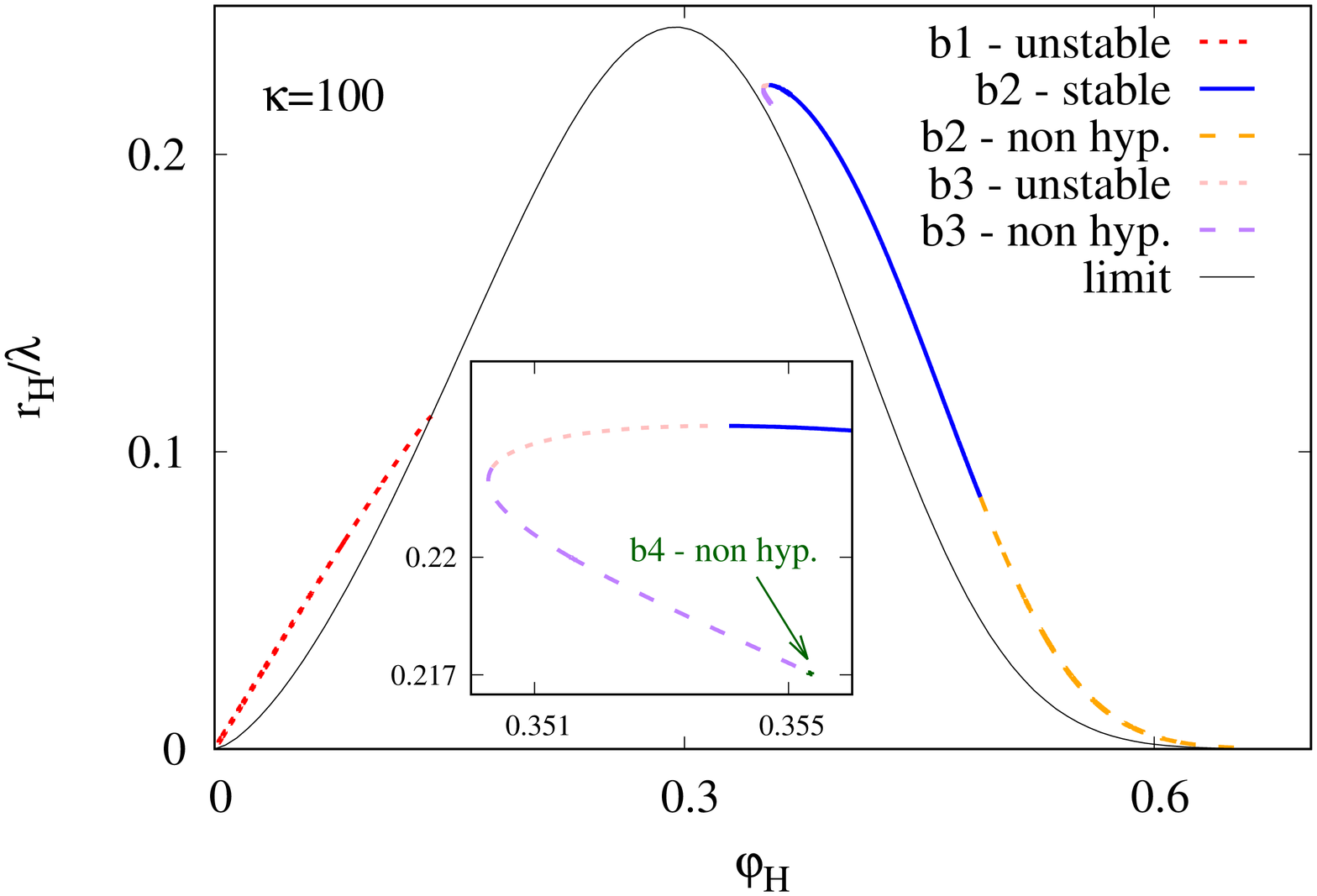}
		\includegraphics[width=0.34\textwidth,angle=-90]{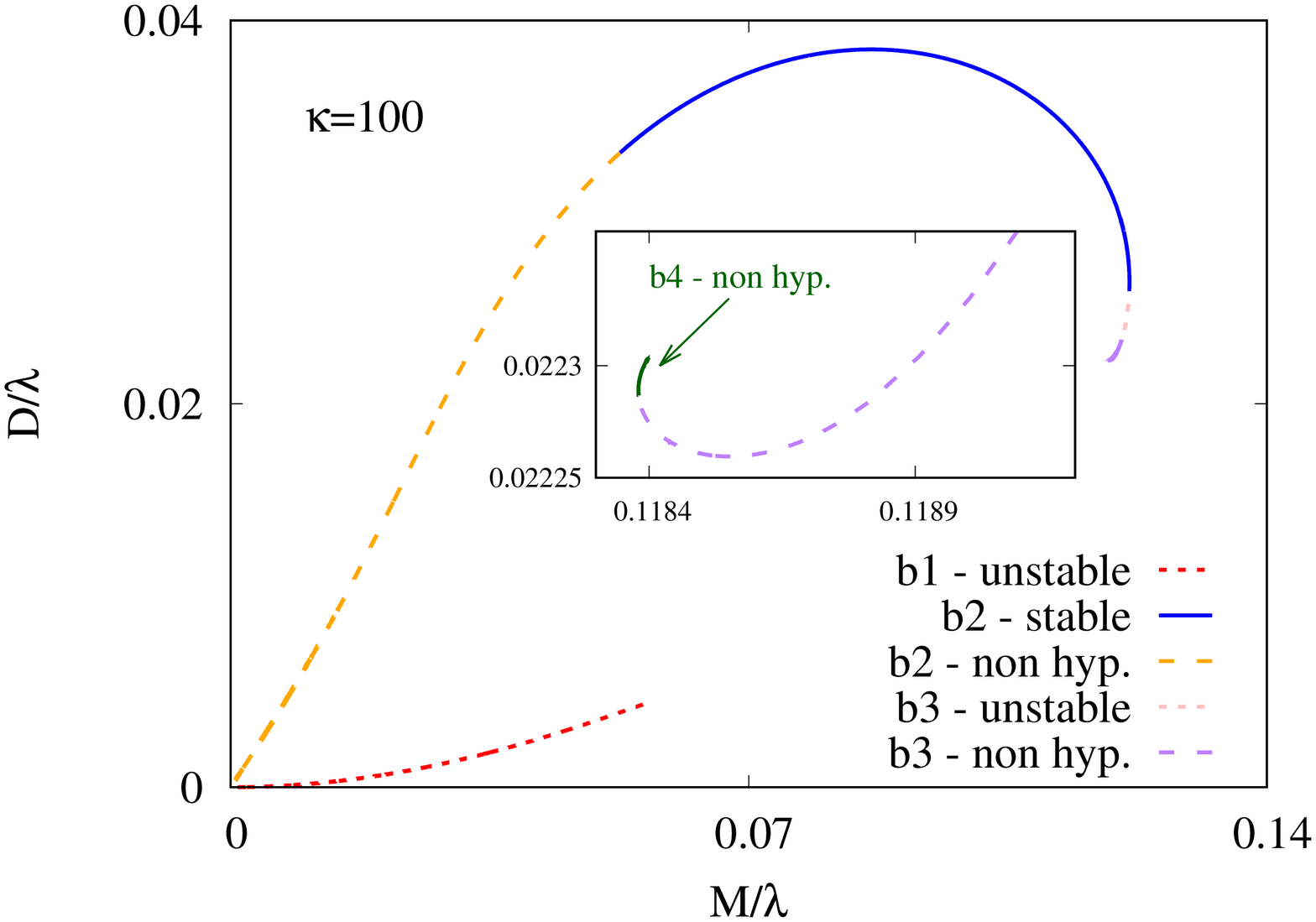}
		\includegraphics[width=0.36\textwidth,angle=-90]{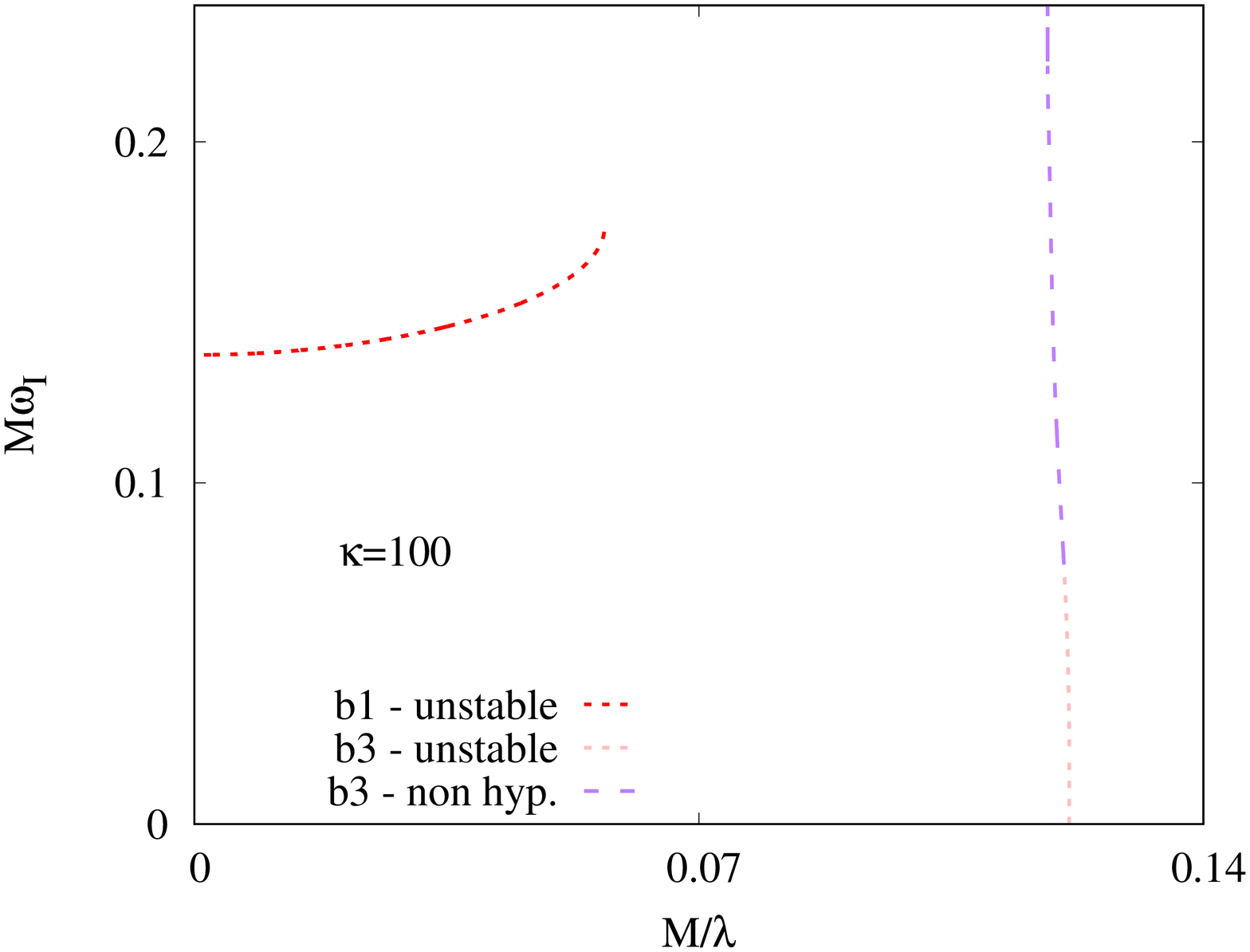}
		\caption{The same type of plots as in Fig. \ref{Fig:k1000_I} but for $\kappa=100$ and coupling $f_I$.}
		\label{Fig:k100_I}
	\end{figure}
	
	Now let us turn to the case of intermediate values of $\kappa$ that is perhaps the most complicated of all. The behavior of the solutions as well as their stability is represented in figure \ref{Fig:k100_I}. Let us first focus on the top-left panel of the figure that shows $r_H/\lambda$ vs $\varphi_H$. In red the solutions from branch b1 are represented that  now end in the limit curve. In blue and in orange we show configurations from branch b2. However, a new branch b3 is found in this case, and it is represented by the pink and purple curves. Configurations in b3 are continuously connected with the solutions in b2. However, branch b3 does not stop at the singular limit curve (solid black line). Instead it is smoothly connected to a new branch of solutions that we label by b4 (green curve).
	As a matter of fact we suspect that more branches like this exist, connecting with b4. If we look at the top-right diagram where $D/\lambda$ is represented as a function of $M/\lambda$, 
	we can see how these branches inspiral around a finite value of the mass and scalar charge, each branch being considerably shorter than the previous one. 
	This is a very interesting phenomenon observed also for other hairy black hole solutions \cite{Herdeiro:2014goa,Kunz:2019bhm} (or for black strings \cite{Kleihaus:2006ee,Kalisch:2016fkm,Kalisch:2017bin}). 
	It is, though, notoriously difficult to obtain these branches from a numerical point of view. As we will see later, they are also unstable {or non-hyperbolic} and that is why we have limited our calculations up to b4.

	Similar to the large $\kappa$ case, solutions in b1 are always unstable but always hyperbolic.We show the unstable mode in
	the bottom panel of figure \ref{Fig:k100_I}, red curve. For the configuration in b1 that ends on the critical solution, the mode has a finite value.
	
	Solutions in b3 are also unstable and possess an unstable mode (pink and purple line). The mode becomes a zero mode at the solution with maximum mass. Interestingly, part of the b3 branch also loses hyperbolicity (purple curve) as we approach the configuration with minimum mass. 
	In fact, as branch b4 connects with b3, it remains non-hyperbolic as it inspirals, and 
	we expect that the additional branches inspiraling after b4 
	will be also either unstable or non-hyperbolic.
	
	Solutions in b2 are again stable (blue), but similar to the large $\kappa$ case, the radial perturbations become non-hyperbolic (orange curve)  for small enough values of $M/\lambda$. With the further decrease of $\kappa$ the part of the branch that is hyperbolic shrinks until it completely disappears making the full b2 branch non-hyperbolic as we will see in the next subsection.

	\subsubsection{Coupling  $f_{I}$, small $\kappa=25$}
	
	\begin{figure}[h]
		\centering
		\includegraphics[width=0.35\textwidth,angle=-90]{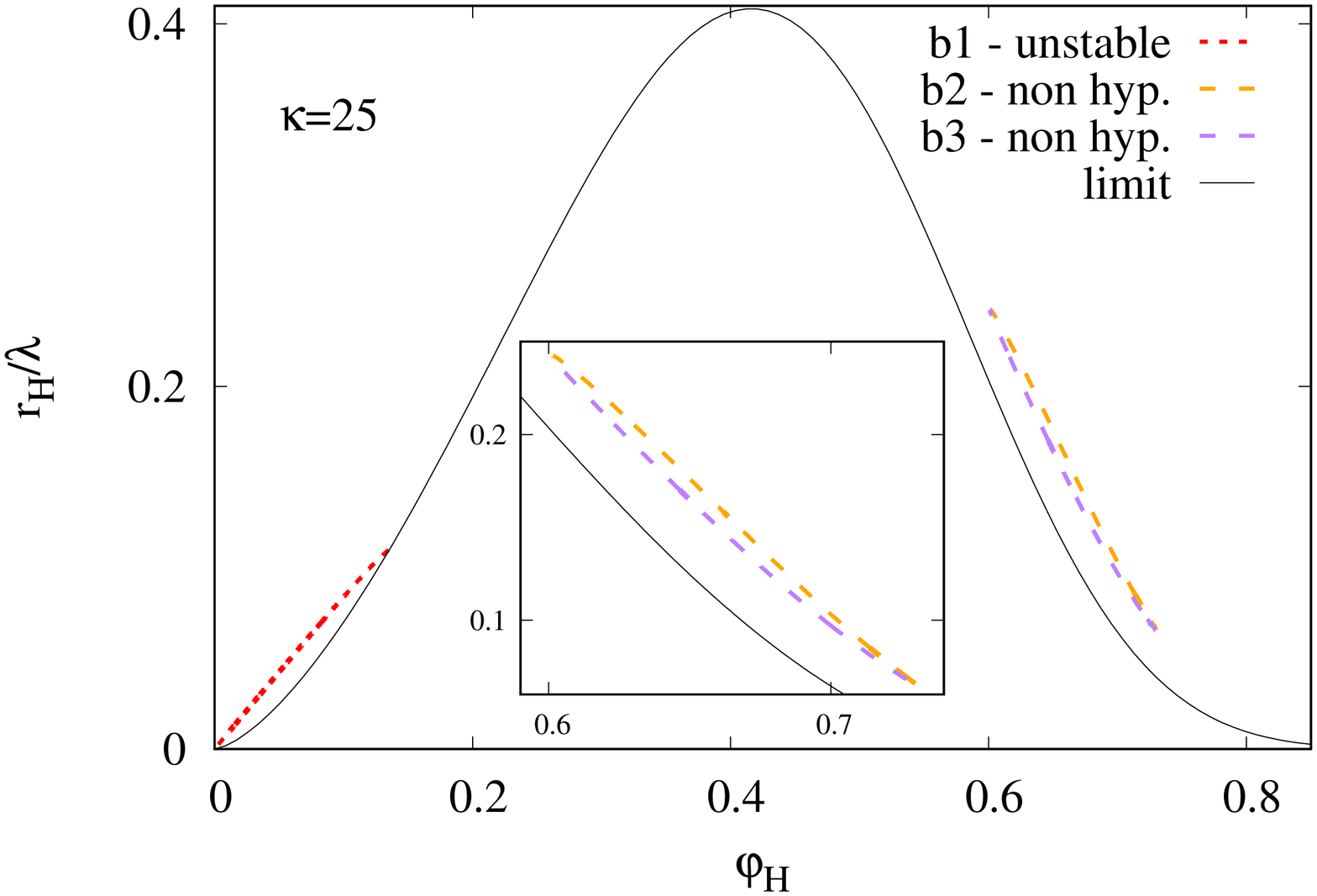}
		\includegraphics[width=0.34\textwidth,angle=-90]{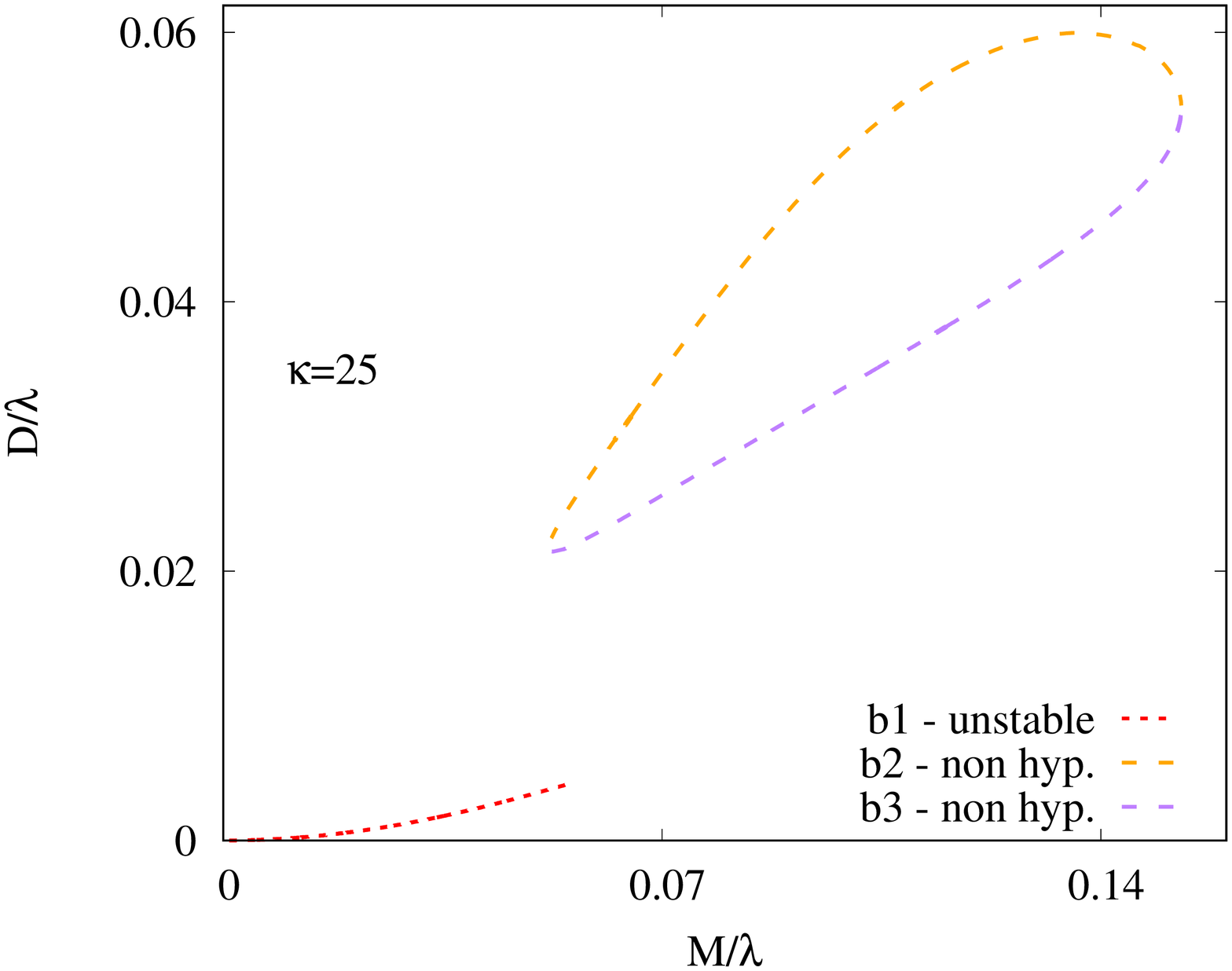}
		\includegraphics[width=0.36\textwidth,angle=-90]{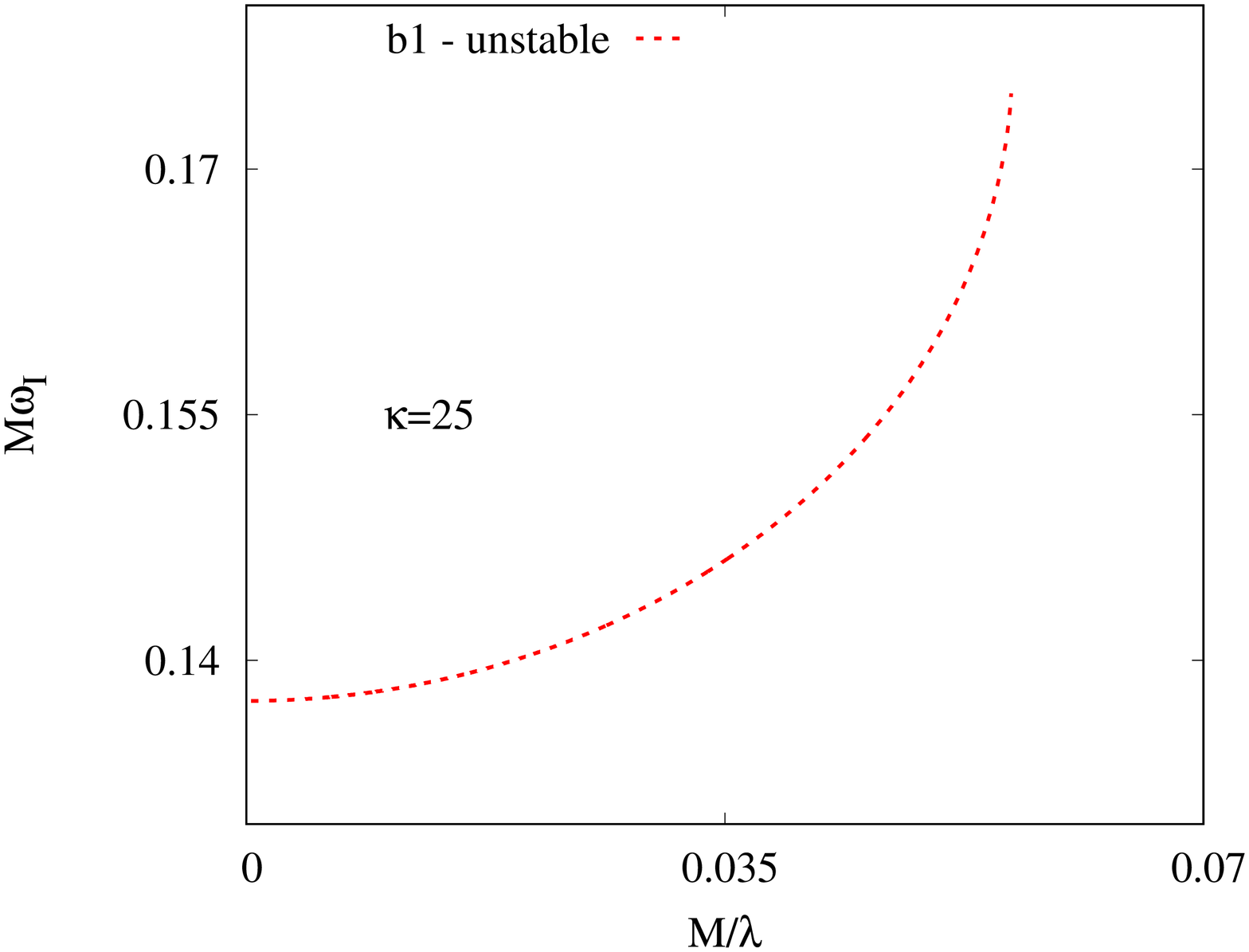}
		\caption{The same type of plots as in Fig. \ref{Fig:k1000_I} but for  $\kappa=25$ and coupling $f_I$.}
		\label{Fig:k25_I}
	\end{figure}
	
	In figure \ref{Fig:k25_I} the results for coupling $f_I$ and small $\kappa=25$ are shown. The spectrum of solutions  in this case is somewhat  simpler. Let us fist look in detail in the top-left panel of the figure where $r_H/\lambda$ vs $\varphi_H$ is plotted. In red we show solutions from branch b1, with similar properties for all $\kappa$ ranges. In orange we have the solutions from b2, which are continuously connected with the b3 solutions (purple). In the top-right panel of \ref{Fig:k25_I} the $D/\lambda$ vs $M/\lambda$ dependence is represented. Branch b2 and branch b3 form a closed loop that lies between two solutions with minimum and maximum mass.
	
	Now let us look at the stability of the solutions that can be concluded from the bottom panel of the figure where the imaginary part of the unstable modes are represented. Solutions in b1 (red curve) are always unstable and hyperbolic, as it can be seen from the bottom panel in the figure, where the unstable mode frequency is plotted for b1. Solutions in b2 (orange) and b3 (purple) are always non-hyperbolic.

	\subsection{Results for coupling  $f_{II}(\varphi)=  \frac{1}{6\kappa}\left(1-e^{-\kappa\varphi^6}\right)$}
	
	\begin{figure}[H]
		\centering
		\includegraphics[width=0.38\textwidth,angle=-90]{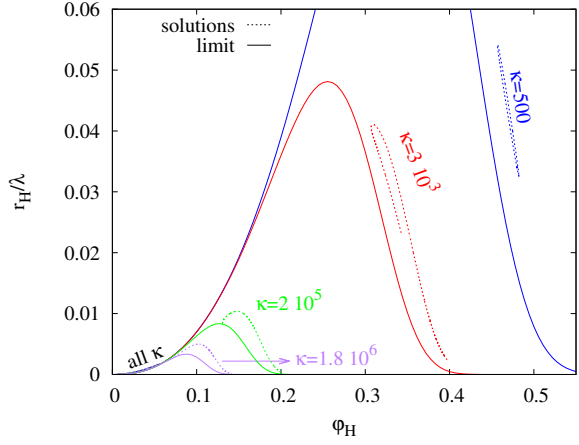}
		\caption{The normalized horizon radius $r_H/\lambda$ as a function of the scalar field at the horizon for $f_{II}$, $\beta=6$ and several $\kappa$. Dashed lines correspond to black hole solutions while the solid line represents the singular limit curve.}
		\label{Fig:results_II}
	\end{figure}
	
	Here we analyze the coupling $f_{II}$ that contains a sextic scalar field term in the exponent instead of the quartic term in $f_I$. It is interesting because it also allows for the existence of scalarized phases. Essentially, the results are similar to the first coupling $f_I$ and that is why we will present them here quickly. In figure \ref{Fig:results_II} we show $r_H/\lambda$ vs $\varphi_H$ for $f_{II}$ and $\kappa=500,3\times 10^3,2\times 10^5,1.8\times 10^6$ in blue, red, green and purple, respectively. The solid lines correspond to the singular limit curves, while the dashed lines represent all scalarized solutions. The general structure of the space of solutions is very similar to the one obtained for $f_I$ where three distinct types of cases, depending on the value of $\kappa$, can be recognized: For large values of $\kappa$, all solutions are smoothly connected. For intermediate values, the limit curve separates the solutions in two parts to the left and to the right. For large values of $\kappa$, the branches on the right side close forming a loop.

	\begin{figure}[H]
		\centering
		\includegraphics[width=0.38\textwidth,angle=-90]{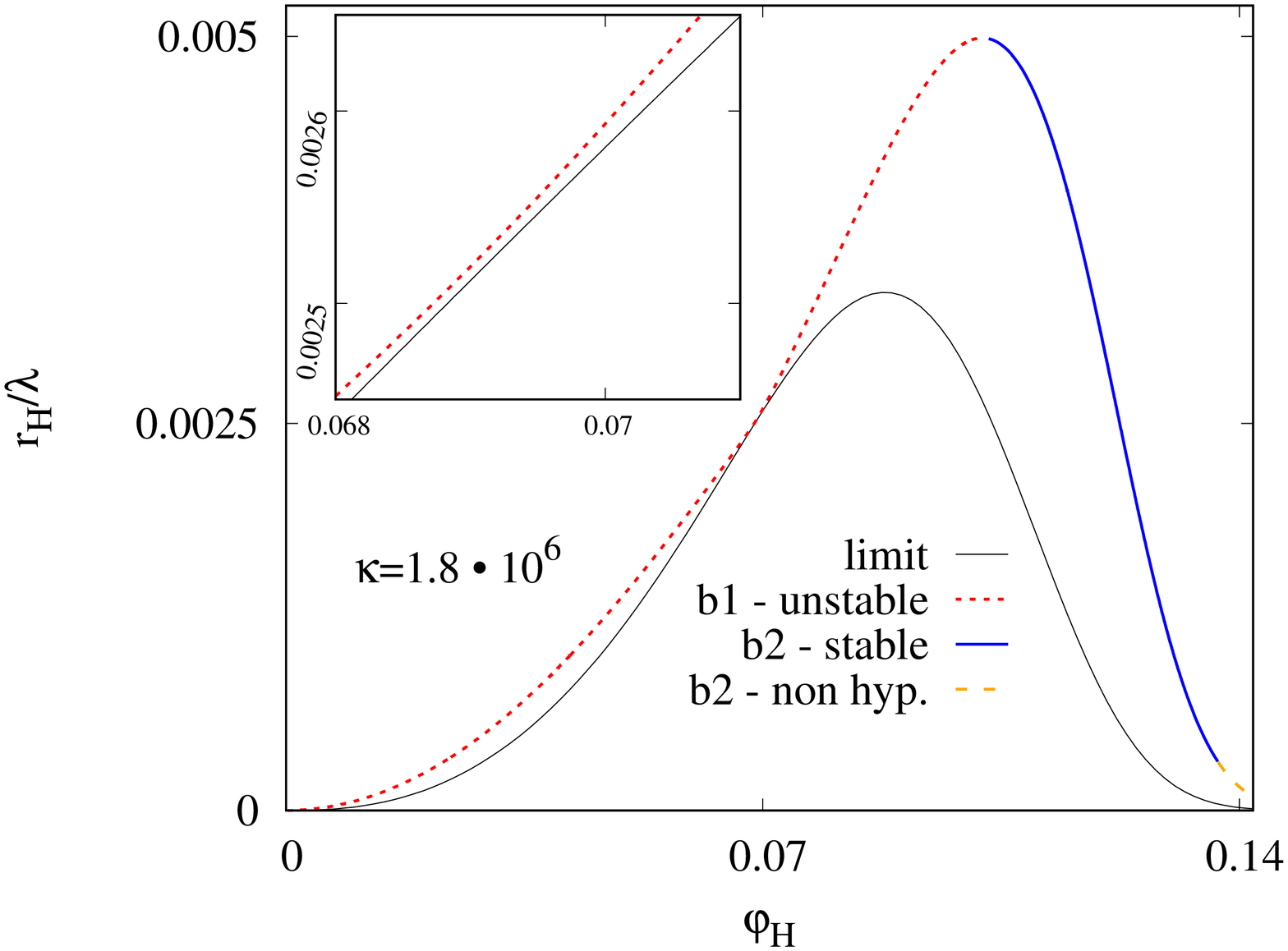}
		\includegraphics[width=0.38\textwidth,angle=-90]{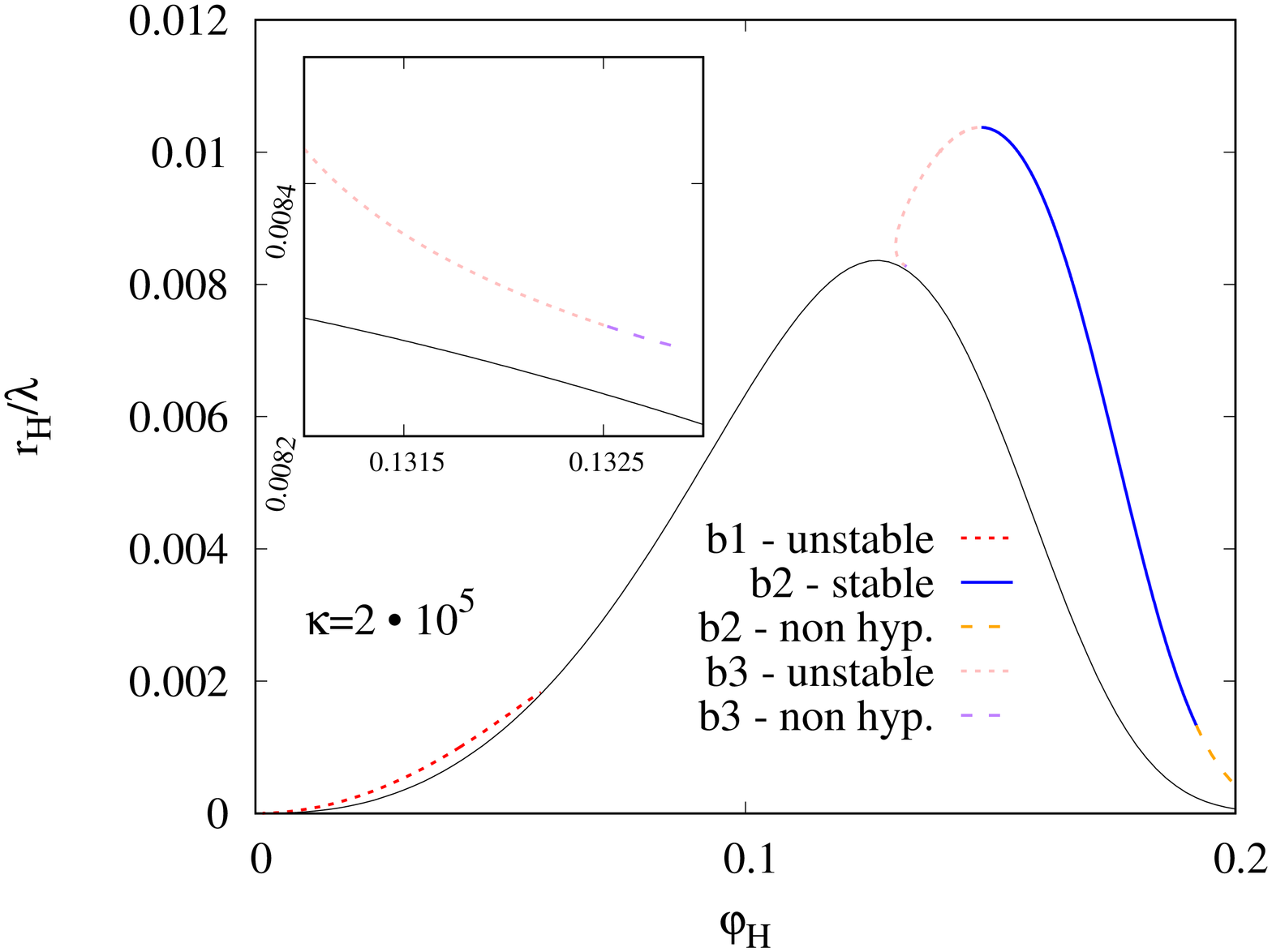}
		\includegraphics[width=0.38\textwidth,angle=-90]{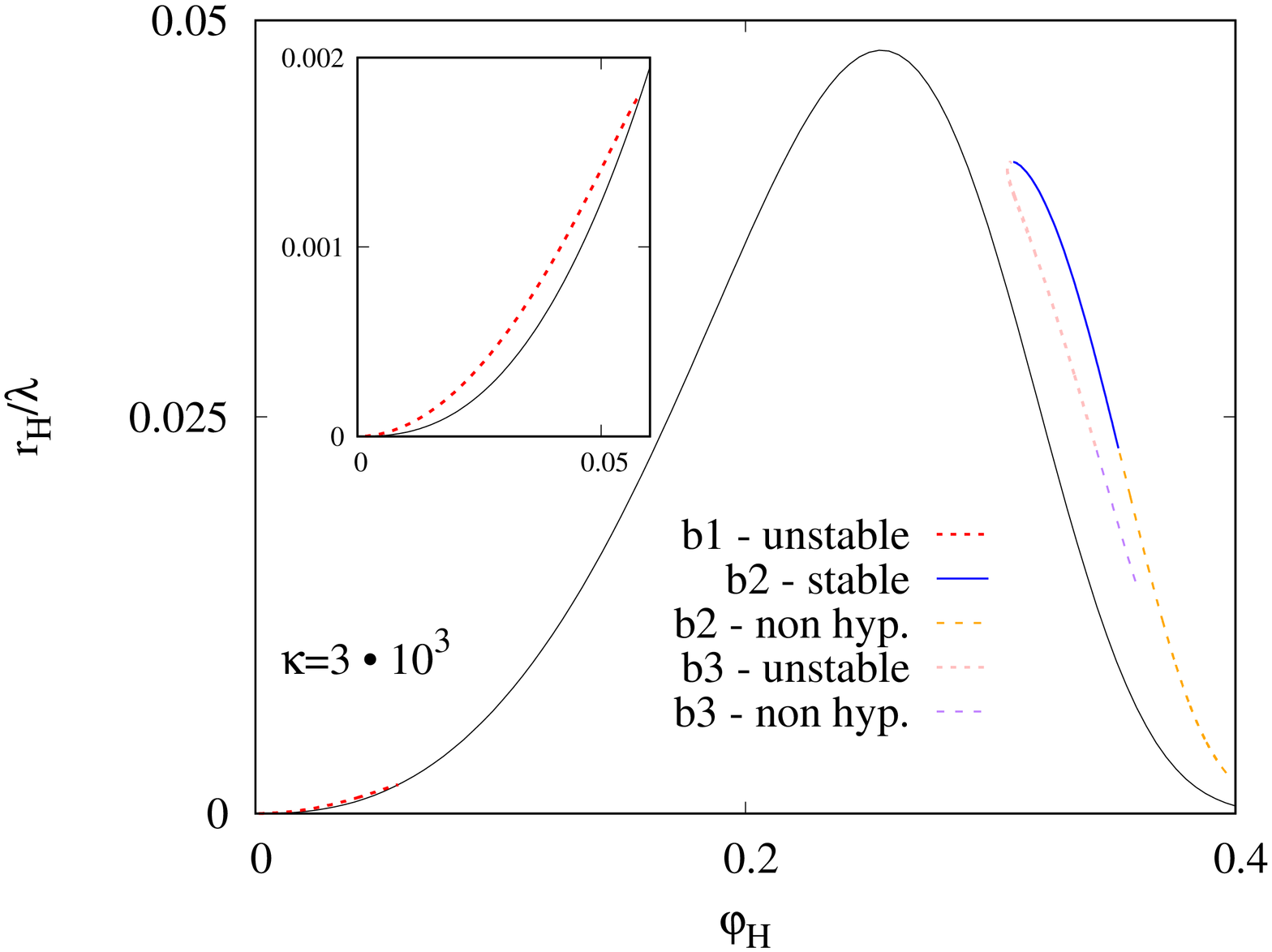}
		\includegraphics[width=0.38\textwidth,angle=-90]{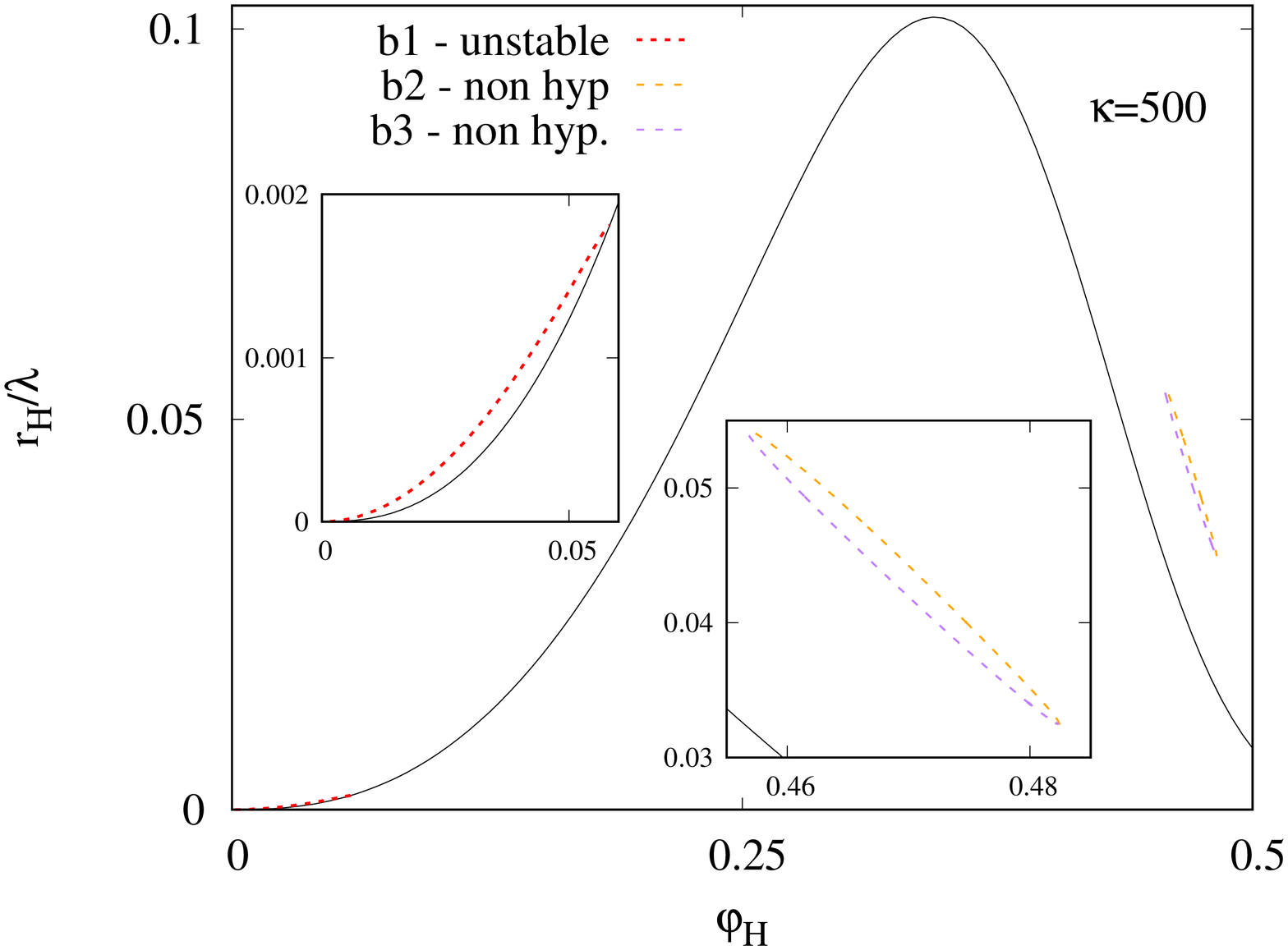}
		\caption{The horizon radius  $r_H/\lambda$ as a function of $\varphi_H$ for $f_{II}$ and $\kappa=1.8\times10^6$ (upper-left), $\kappa=2\times10^5$ (upper-right), $\kappa=3000$ (bottom-left), and $\kappa=500$ (bottom-right). 
			The solid black line represents the singular limit curve while the lines with different colors correspond to different branches of hairy black hole.		
			The stability and hyperbolicity of the branches are explicitly marked in the figure. 
		}
		\label{Fig:k1.8e6_II}
	\end{figure}

	Since the stability properties of the branches in this case are qualitatively very similar to the $f_I$ coupling, we will present the results for all interesting ranges of $\kappa$ in the single figure \ref{Fig:k1.8e6_II}. {There we show $r_H/\lambda$ vs $\varphi_H$, and there is a single plot for each interesting $\kappa$ range with a representative value of $\kappa$ chosen. The stability and hyperbolicity of the branches are explicitly marked.} As one can see the b1 branches are all unstable. All b3 branches in the figures are also either unstable or non-hyperbolic. The only branch that can be stable is b2. As one can see, this branch also loses hyperbolicity for small black hole masses and the non-hyperbolic region gets bigger with the decrease of $\kappa$. 
	The stable hyperbolic part of the branch completely disappears for small enough $\kappa$, as can be observed in the bottom-right panel of Fig.~\ref{Fig:k1.8e6_II}. 
	On the other hand, for intermediate values of $\kappa$, we also expect b3 to be smoothly connected to branches that inspiral around some particular values of the black hole parameters (see for instance the inset in the top right figure). But these branches will most likely be unstable or non-hyperbolic.

	\subsection{Results for coupling function $f_{III}(\varphi)=  \frac{1}{2\beta}\left(1-e^{-\beta (\varphi^2+\kappa\varphi^4)}\right)$}

	\begin{figure}[H]
		\centering
		\includegraphics[width=0.38\textwidth,angle=-90]{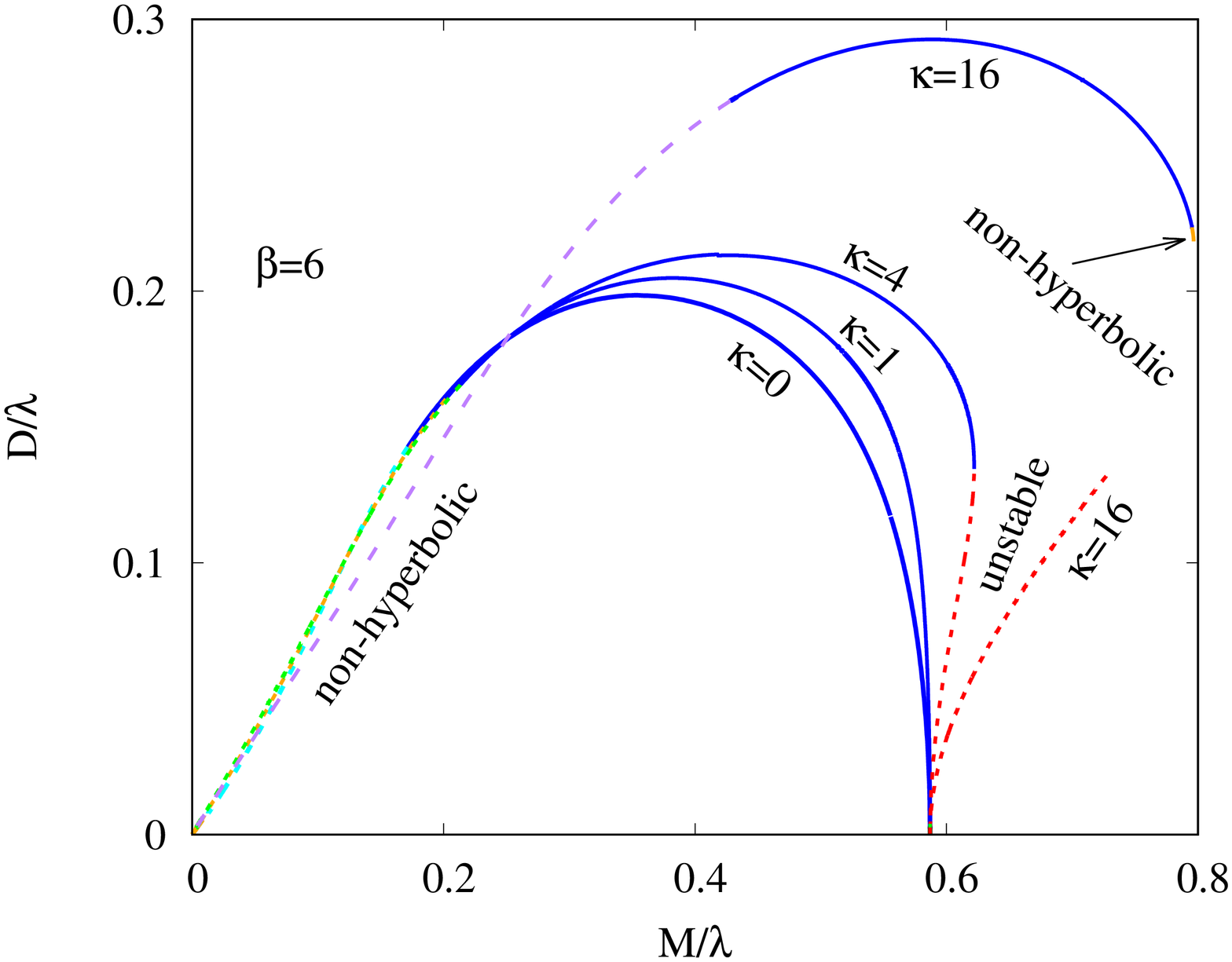}
		\includegraphics[width=0.38\textwidth,angle=-90]{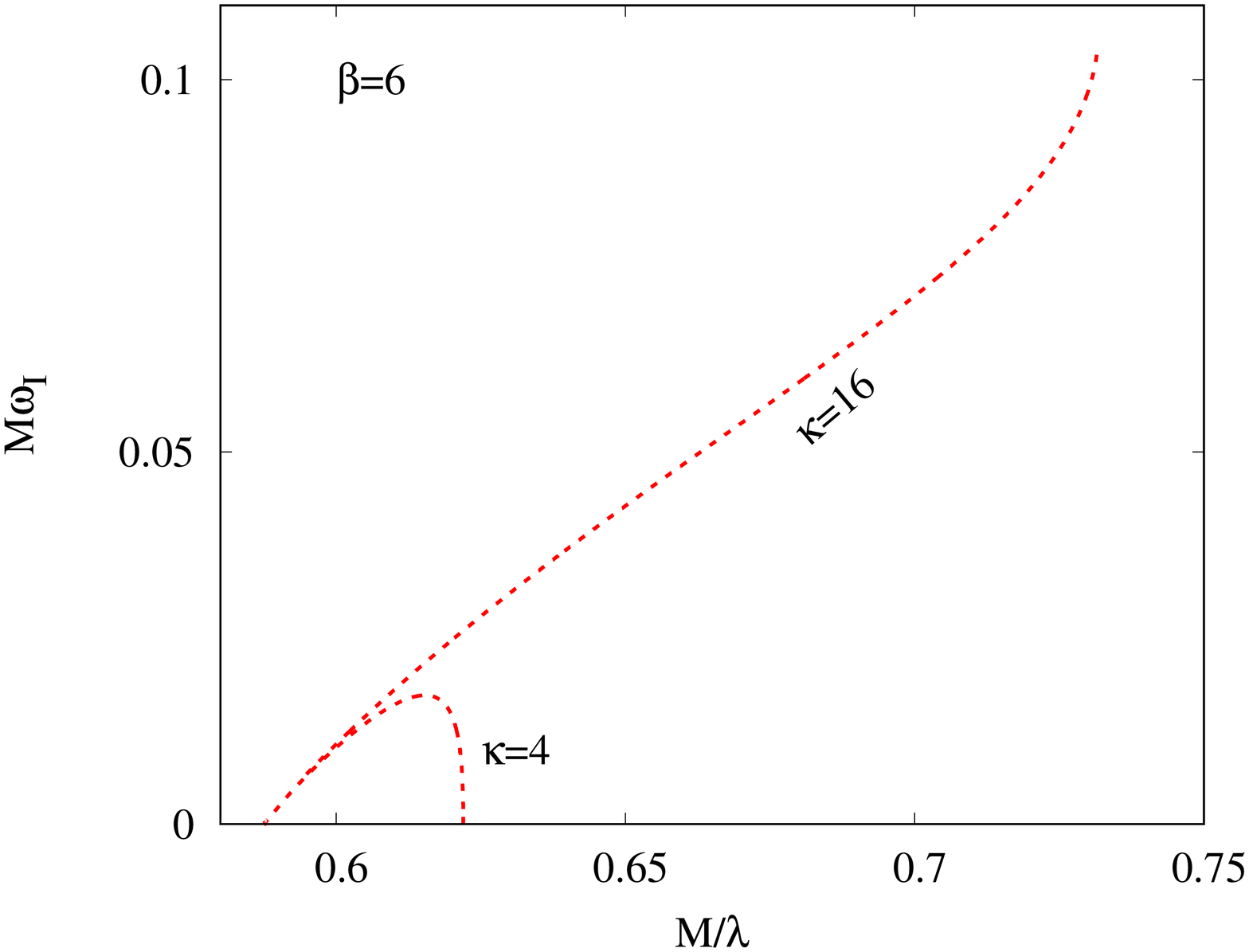}
		\caption{Results for the scalarized black hole branches and their stability for the third coupling $f_{III}$. \textit{(left panel)} The normalized scalar charge $D/\lambda$ as a function of $M/\lambda$. \textit{(right panel)} The imaginary part of the frequency normalized with the black holes mass $M \omega_I$ for the unstable branches in the left panel, as a function of $M/\lambda$.}
		\label{Fig:results_III}
	\end{figure}
	
	Here we will analyze the $f_{III}$ coupling function that allows for the presence of both the standard scalarization and the nonlinear scalarized phases. The structure of the solution branches in this case is simpler compared to $f_I$ and $f_{II}$ and it is presented in Fig.~\ref{Fig:results_III}. We fix $\beta=6$, meaning that the quadratic part of the coupling is always on, and only $\kappa$ is varied. This parameter is, loosely speaking, responsible for the appearance and the behavior of the nonlinear scalarized phases. We consider {four values of $\kappa=0,1,4,16$}, meaning we deform the quadratic coupling.
	
	In figure \ref{Fig:results_III} (left) we show $D/\lambda$ vs $M/\lambda$. For $\kappa=0$ we have the well known curve of standard scalarized solutions: these black holes possess a non-trivial scalar field with no nodes. They branch from the Schwarzschild solution at the point where it becomes unstable and extend down to $M/\lambda=0$. These black holes are radially stable, but for small values of the mass, they lose hyperbolicity, as indicated in the plot. When $\kappa=1$, the contribution from the $\phi^4$ coupling is still small, and the qualitative behavior of the standard scalarized branch is the same as the $\kappa=0$ case. We should note that in the figure we have plotted only the so-called fundamental branch of solutions that have no nodes of the scalar field. More branches of solutions exist for smaller masses with increasing number of scalar field nodes, but all of them are unstable \cite{Blazquez-Salcedo:2018jnn} and thus we will not pay special attention to them.
	
	With the increase of $\kappa$, though, the picture gradually changes and already for $\kappa=4$ the deviations due to the quadratic coupling are more significant. After the bifurcation point the scalarized branch turns right instead of left, which is normally a signal of instability observed also in the case of pure standard scalarization with $\kappa=0$ but for smaller $\beta$ \cite{Doneva:2017bvd,Doneva:2018rou,Silva:2018qhn}. Thus we expect that here as well part of the fundamental branch is no longer stable. In fact, uniqueness on this branch with respect to the total mass is lost: It is possible to find two node-less solutions with the same mass and different scalar charges. However the solution with lower scalar charge is unstable up to the maximum of the mass. This can be seen in figure \ref{Fig:results_III} (right), where we plot the unstable radial mode for the solutions.   
	
	Note that for $\kappa=4$ there are two zero modes, corresponding to the two branching solutions: Schwarzschild ($M/\lambda=0.587$) and the maximum mass configuration ($M/\lambda=0.622$). {From this maximum mass configuration emerges the second branch that turns out to be radially stable. Thus, our interpretation is that  we can formally divide the scalarized solution curve in two branches before and after the maximum of the mass, similar to the other coupling functions $f_I$ and $f_{II}$. The red curve would correspond to the branch b1 while the blue and orange correspond to b2.}  Similar to what we find for $\kappa=0,1$ though, these solutions become non-hyperbolic for small enough values of the mass.

    Apart from the visual analogy with the other coupling functions, we have also a physical motivation for associating the larger $D/\lambda$ branch with {b2}. The reason is that it is stable but unconnected to Schwarzschild (at least not through a sequence of stable solutions). Part of it also spans a region where the Schwarzschild solution is stable. In order to ``jump'' from the GR branch to the one with stable hairy black holes, a large initial perturbation will be required that is exactly the mechanism for the formation of black hole scalarized phases described above. 

	In figure \ref{Fig:results_III} (left) we also show solutoions with $\kappa=16$. In this case the limit curve separates the solutions in to two different
	sets. After the bifurcation point, the fundamental scalarized branch turns right, and it is possible to obtain an unstable mode along this branch (see figure \ref{Fig:results_III} (right)). The stable branch is no longer connected with the unstable solutions, but instead is surrounded by non-hyperbolic branches, very similar to what we obtain for the couplings $f_I$ and $f_{II}$ for intermediate values of $\kappa$. This is another motivation to associate the larger $D/\lambda$ branch with b2.

	\section{Conclusions}
	In the present paper we have studied the radial stability and hyperbolicity of {the so-called nonlinear scalarized phases of Schwarzschild black holes within the beyond spontaneous scalarization scenario}. These are black hole solutions with nontrivial scalar field that co-exist with the (always) linearly stable Schwarzschild solutions and can be excited only nonlinearly if a strong enough perturbation is imposed. The spectrum of these solutions can be very complicated possessing several distinct branches and it is interesting to identify the stable solutions that are potentially relevant for astrophysics. 
	
	Depending on the exact choice of the coupling function between the scalar field and the Gauss-Bonnet invariant, the number and the structure of hairy black hole branches can vary significantly. In some cases these branches are smoothly connected to each other and in others they are terminated because they start violating the regularity condition at the black hole horizon. For certain values of the parameters $\kappa$ in the coupling functions \eqref{eq:coupling_function_I} and \eqref{eq:coupling_function_II} we have even found a series of inspiraling branches that is a new result unobserved before in sGB gravity. Despite the complicated structure of solutions it turns out that only one scalarized branch can be stable, and that coincides with the results from the thermodynamical analysis in \cite{Doneva:2021tvn}. This branch either spans from zero black hole mass to a maximum mass, as happens for large $\kappa$, or it is limited between two finite nonzero values of the mass for small $\kappa$. As far as hyperbolicity is concerned it turns out that it is lost for small black hole masses in the former case, while in the latter we can easily have a situation where the whole {potentially stable} branch consists of non-hyperbolic solutions with respect to the radial perturbation equation. This effectively means that for certain coupling parameters it is not possible to have both stable and hyperbolic scalarized phases of the Schwarzschild black hole. We should point out, though, that this undesired behavior happens only in a limited range of small $\kappa$ while for the majority of cases we can have stable and hyperbolic black hole solutions that are viable astrophysical candidates.
	
	We have focused our analysis also on a second type of coupling function given by \eqref{eq:coupling_function_III} that can be viewed as a connection between the standard scalarization and the nonlinear scalarized phases. Namely, the Schwarzschild black hole still becomes unstable below a certain mass, but the standard scalarized black holes branching out at this point are unstable and smoothly connected to a new branch of solutions whose properties resemble closely the scalarized phases. More specifically, this new branch also exists for values of the parameters where the Schwarzschild black hole is stable and solutions can form scalar hair only if a sufficiently large nonlinear perturbation is imposed. The radial stability analysis indeed shows that this new branch of solutions is stable and, similar to the case of standard scalarization, it loses hyperbolicity for small black hole masses. 
	
	All these results place the nonlinear scalarized phases of the Schwarzschild black hole on a solid basis, showing that they are both stable and hyperbolic for a sufficiently large range of parameters. The fact that they are not smoothly connected with the stable Schwarzschild black hole offers a variety of astrophysical implications. Namely, the transition from one to the other will happen with a jump that will leave a distinct gravitational wave and electromagnetic signature. In order to study such processes in detail, though, we need to examine the nonlinear dynamics of the system that will be the topic of  future work.

	\bibliographystyle{unsrt}

\end{document}